\documentclass{PoS}
\usepackage{amsmath}
\newcommand{\braket}[1]{\langle#1\rangle}
\newcommand{\qq}{\braket{qq}}

\newlength{\colw}
\newcommand{\cphases}{\cite{Cotter:2012mb}}
\newcommand{\ctrans}{\cite{Boz:2013rca}}
\newcommand{\cprev}{\cite{Hands:2006ve,Hands:2010gd,Cotter:2012mb,Boz:2013rca}}
\newcommand{\ccurr}{\cite{Cotter:2012mb,Boz:2013rca}}

\title{Phase transitions in dense 2-colour QCD}

\ShortTitle{Phase transitions in QC$_2$D}

\author{Tamer Boz, Seamus Cotter, Leonard Fister, \speaker{Jon-Ivar
    Skullerud}\\ 
        Department of Mathematical Physics, National University of
        Ireland Maynooth, Maynooth, County Kildare, Ireland\\
        E-mail: \email{jonivar@thphys.nuim.ie}}

\abstract{We investigate 2-colour QCD with 2 flavours of Wilson
  fermion at nonzero temperature $T$ and quark chemical potential $\mu$,
  with a pion mass of 700 MeV ($m_\pi/m_\rho=0.8$). From temperature scans at
  fixed $\mu$ we find that the critical temperature for the superfluid to
  normal transition depends only very weakly on $\mu$ above the onset
  chemical potential, while the deconfinement crossover temperature is
  clearly decreasing with $\mu$. We
  also present results for the Landau-gauge gluon propagator in the
  hot and dense medium.}

\FullConference{31st International Symposium on Lattice Field Theory - LATTICE 2013\\
		July 29 - August 3, 2013\\
		Mainz, Germany}

\begin{document}

\section{Introduction}

Despite intensive theoretical efforts over the past decade and more,
we do still not have a quantitative understanding of QCD at large
baryon density.  This is primarily due to the sign problem preventing
first-principles Monte Carlo simulations in this r\'egime. 
One way of circumventing this
is to study QCD-like theories without a sign problem, and use
these to provide a benchmark for model studies and other methods which
do not suffer from the sign problem.  The simplest such theory, which
shares with QCD the properties of confinement and dynamical symmetry
breaking, is 2-colour QCD (QC$_2$D).

In a series of papers \cprev\ we have studied QC$_2$D with 2 flavours
of Wilson fermion at nonzero baryon chemical potential $\mu$ and
temperature $T$, culminating in a tentative mapping out of the phase
diagram in the $(\mu,T)$ plane \cite{Cotter:2012mb,Boz:2013rca}.  Here
we will report on the determination of the phase transition lines
\ctrans\ and present new results for the gluon propagator at nonzero
$\mu$ and $T$.  Updated results for the equation of state are
presented in a separate talk \cite{Cotter:2013lat}.


We use a standard Wilson gauge and fermion action augmented with a
diquark source term to lift low-lying eigenvalues in the superfluid
phase.  The lattice spacing is $a=0.178(6)$fm and $m_\pi/m_\rho$=0.8,
with $am_\pi=0.645(8)$ \cphases.  We have performed simulations at
four fixed temperatures, $T=47, 70, 94$ and 141 MeV, corresponding to
$N_\tau=24, 16, 12$ and 8 respectively, for a range of chemical
potentials $\mu a=$0.0--0.9.  At $\mu a=0.35, 0.4, 0.5$ and 0.6 we
have also performed temperature scans on $16^3\times N_\tau$ lattices
with $N_\tau=$4--16.  For the diquark source $j$ we have used
$ja=0.02, 0.04$ in order to allow an extrapolation to the physical
$j=0$ limit.  We refer to \ccurr\ for further details about the action
and parameters.

\section{Superfluid to normal transition}

\begin{figure}[tb]
\includegraphics*[width=\textwidth]{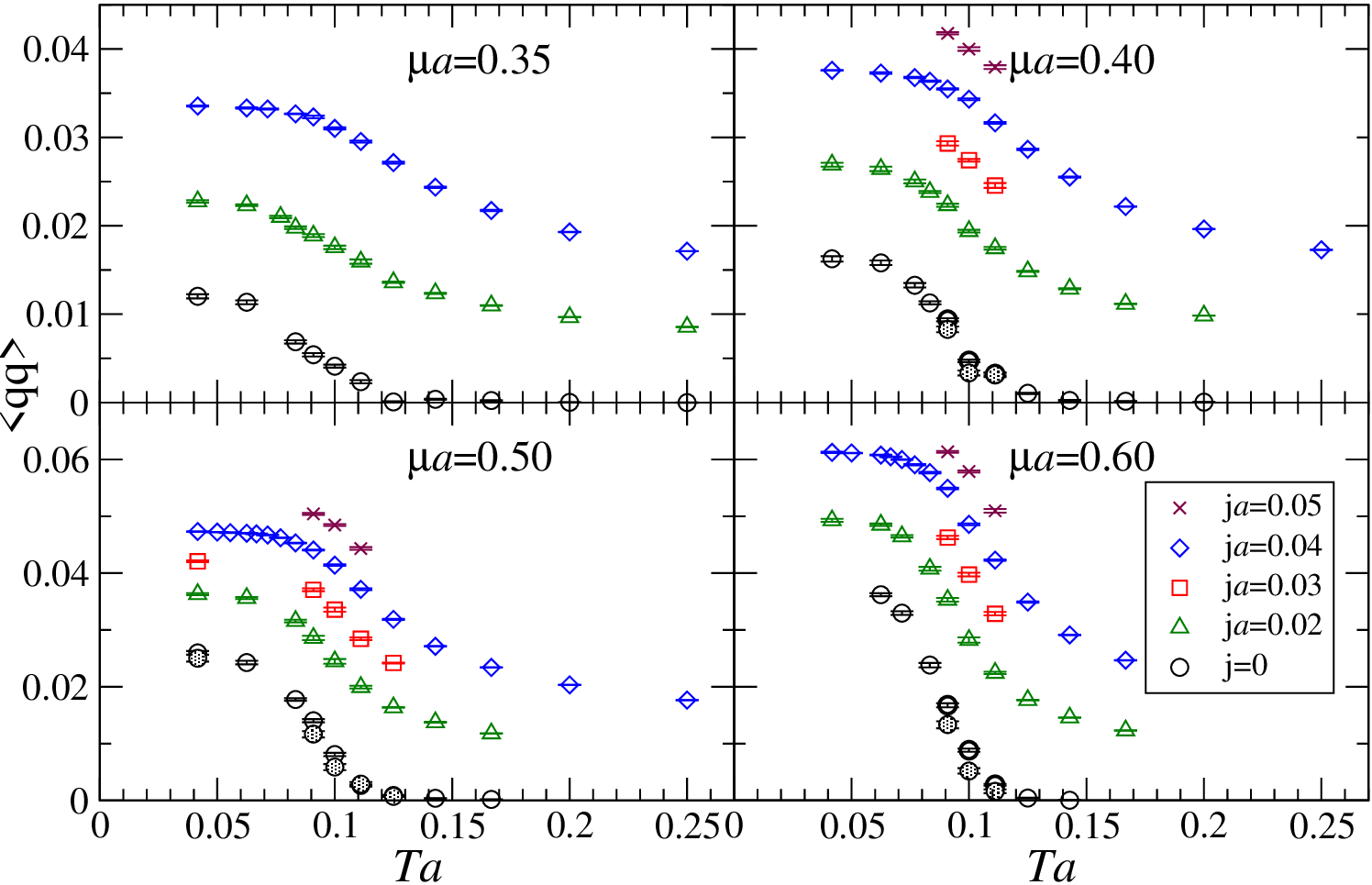}
\caption{Diquark condensate $\qq$ as a function of temperature $T$ for
  chemical potential $\mu a=0.35, 0.4, 0.5, 0.6$ (top to bottom).  The
circles are data extrapolated to $j=0$ using a linear Ansatz for
$ja\leq0.04$; the shaded circles denote the results of a linear
extrapolation using $ja=0.02,0.03$ only.}
\label{fig:diquark}
\end{figure}

Figure~\ref{fig:diquark} shows the order parameter for superfluidity,
the (unrenormalised) diquark condensate $\qq$,
as a function of the temperature $T$, for $\mu a=0.35, 0.4,0.5$ and 0.6.
Also shown are the results of a linear extrapolation to $j=0$.  We can
clearly observe a transition from a superfluid phase, characterised by
$\qq\neq0$, at low temperature, to a normal phase with $\qq=0$ at high
temperature, with a transition in the region $0.08\lesssim
Ta\lesssim0.12$ for all four values of $\mu$.
%

We have estimated the critical temperatures $T_s$
for the superfluid to normal transition by determining the inflection
points for $\qq$ at $ja=0.02$ and 0.04, and extrapolated the resulting
values to $j=0$ using a linear Ansatz.  The results are shown in fig.~\ref{fig:phasediag}.
We see that $T_s$ is remarkably constant over the whole range of
$\mu$-values considered.  The indications are that the transition
happens at a somewhat lower temperature at $\mu a=0.35$, but this
point is already very close to the onset from vacuum to superfluid at
$T=0$, $\mu_oa=m_\pi a/2=0.32$, suggesting that $T_s(\mu)$ rises very
rapidly from zero at $\mu=\mu_o$ before suddenly flattening off.

\section{Deconfinement transition}

\label{sec:deconfine}

\begin{figure*}
\includegraphics*[width=\textwidth]{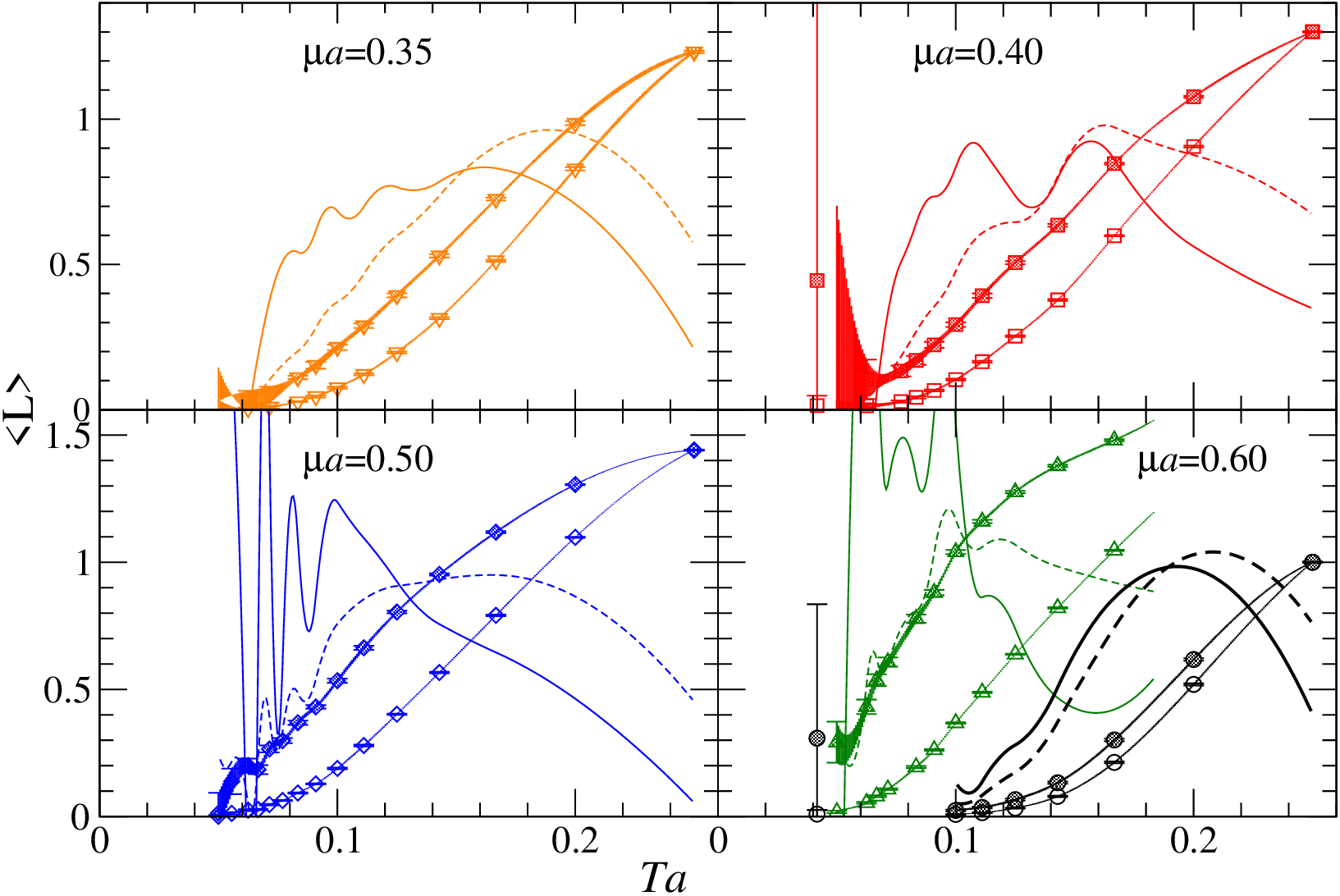}
\caption{The renormalised Polyakov loop $\braket{L}$ as a function of
  temperature $T$ for $ja=0.04$ and $\mu a=0.35,0.4,0.5, 0.6$, with
  two different renormalisation schemes: Scheme A (solid symbols) and
  Scheme B (open symbols), see text for details.  The solid (dashed)
  lines are the derivatives of cubic spline interpolations of the data
  points for Scheme A (B).  The smaller, shaded symbols are results
  for $ja=0.02$.  The black circles and thick lines in the bottom
  right panel are the $\mu=j=0$ results from \cphases.}
\label{fig:polyakov-allmu}

\end{figure*}
The Polyakov loop $\braket{L}$ serves as the traditional order
parameter for deconfinement in gauge theories, with $\braket{L}\neq0$
signalling the transition to a deconfined phase.  Strictly speaking,
$\braket{L}$ is never zero in a theory with dynamical fermions, but it
typically increases with temperature from a very small value in a
fairly narrow region, which may be identified with the deconfinement
transition region.
Unlike the diquark condensate, the renormalisation of the Polyakov
loop depends on temperature; specifically, the relation between
the bare Polyakov loop $L_0$ and the renormalised Polyakov loop $L_R$
is given by
$L_R(T,\mu)= Z_L^{N_\tau}L_0((aN_\tau)^{-1},\mu)$.
In order to investigate the sensitivity of our results to the
renormalisation scheme, we have used two different conditions to
determine the constant $Z_L$ \ctrans, $L_R(T=T_0,\mu=0)=c$, with
$T_0=\frac{1}{4}a^{-1}$ and $c=1$ (Scheme A) or $c=0.5$ (Scheme B).
Figure~\ref{fig:polyakov-allmu} shows $\braket{L}$ evaluated in both
schemes, as a function of temperature.  The Scheme B data have been
multiplied by 2 to ease the comparison with the Scheme A data.  Also
shown are cubic spline interpolations of the data and the derivative
of these interpolations, with solid lines corresponding to Scheme A
and dotted lines to Scheme B.
 
At all $\mu$, we see a transition from a low-temperature confined
region to a high-temperature deconfined region.  In contrast to the
diquark condensate, we see a clear, systematic shift in the transition
region towards lower temperatures as the chemical potential increases.
For all four $\mu$-values, the Polyakov loop shows a nearly linear
rise as a function of temperature in a broad region, suggesting that 
the transition is a smooth crossover rather than a true phase
transition.  This is reinforced by the difference between Scheme A and
Scheme B, with the crossover occuring at higher temperatures in Scheme
B.  At $\mu=0$, the difference between the two schemes is small, but
increases with increasing $\mu$, suggesting a broadening of the
crossover.

Because of the smaller value of $Z_L$, our results for Scheme B are
considerably less noisy than those for Scheme A.  For this reason, we
choose to define the crossover region to be centred on the inflection
point from Scheme B, with a width chosen such that it also encompasses
the onset of the linear region from Scheme A. 

\begin{figure}
\begin{center}
\includegraphics*[width=0.6\textwidth]{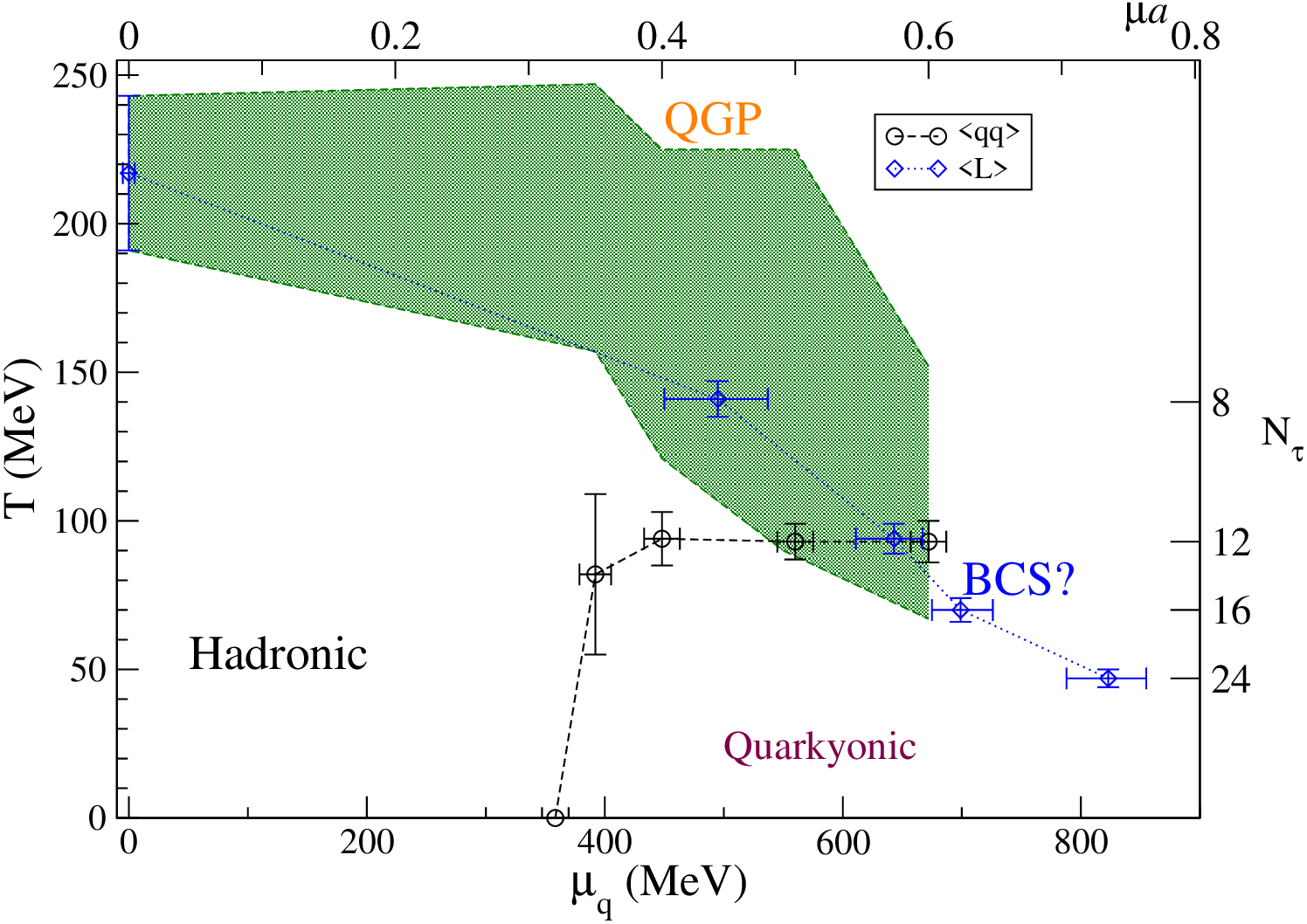}
\end{center}
\caption{Phase diagram of QC$_2$D with $m_\pi/m_\rho=0.8$.  The black
  circles denote the superfluid to normal phase transition; the green
  band the deconfinement crossover.  The blue diamonds are the
  estimates for the deconfinement line from \cphases.}
\label{fig:phasediag}
\end{figure}

The transition region taken from the $ja=0.04$ data
is shown in fig.~\ref{fig:phasediag}.  From
Fig.~\ref{fig:polyakov-allmu} we see that at low $T$, the value of $\braket{L}$
increases as $j$ is reduced, and at $\mu a=0.6$, the
crossover region will most likely move to smaller $T$ in the $j\to0$
limit.  However, we do not have sufficient statistics for $ja=0.02$ at
low $T$ to make any quantitative statement about this.

\section{Gluon propagator}
\label{sec:gluon}

\begin{figure*}
\includegraphics*[width=\textwidth]{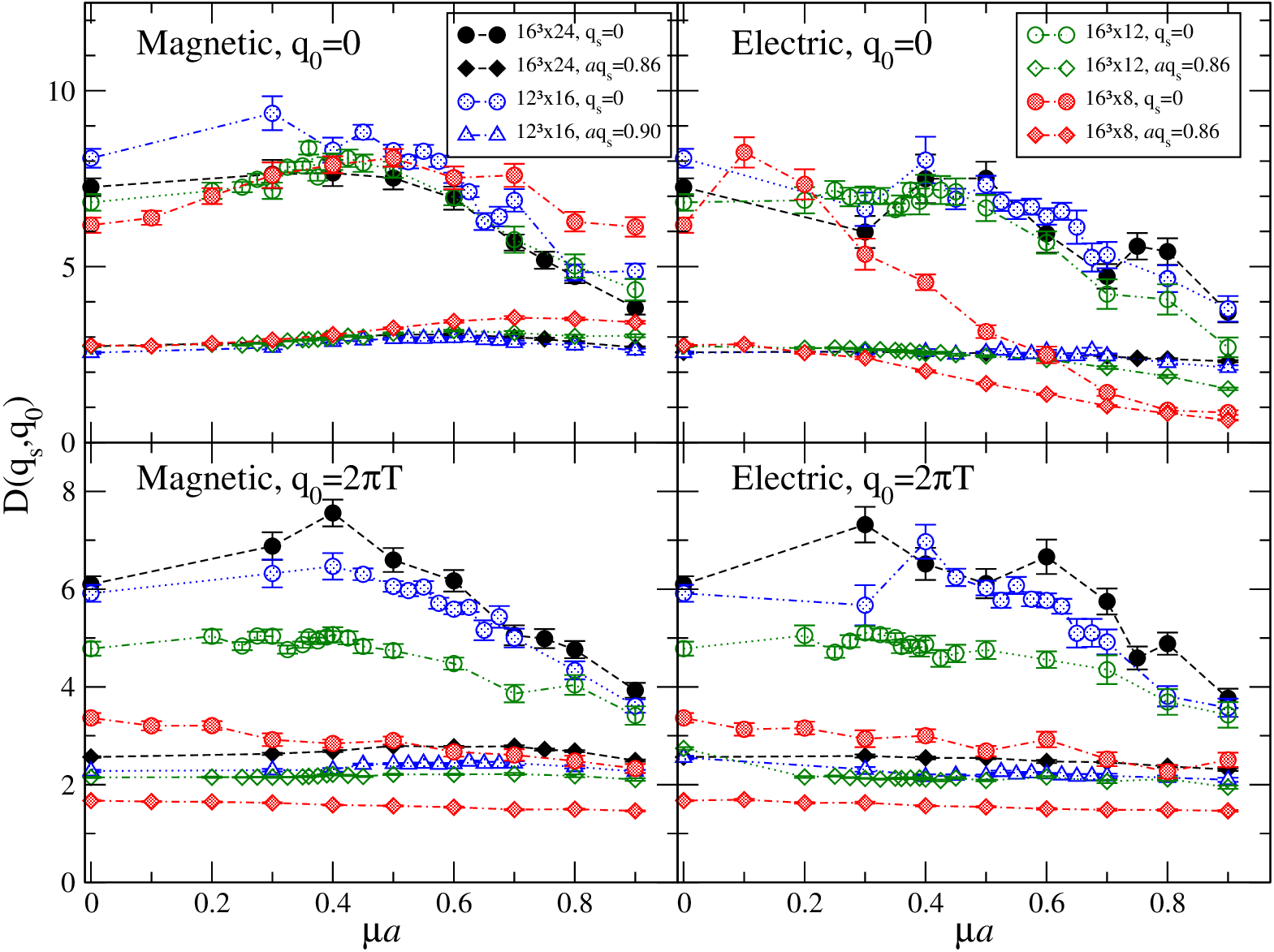}
  \caption{The zeroth (top) and first (bottom) Matsubara mode of the
    magnetic (left) and electric (right) gluon propagator as a
    function of chemical potential $\mu$ for selected values of the
    spatial momentum $q_s=|\vec{q}|$, and different temperatures.}
\label{fig:gluon-compare}  
\end{figure*}

One of the main motivations for studying dense QC$_2$D on the lattice
is to provide constraints on approaches which do not suffer from the
sign problem.  The gluon propagator provides a key input for several
of these approaches, in particular functional studies using the
functional renormalisation group or Dyson--Schwinger equations.  These
are most often carried out in the Landau gauge.

In Landau gauge only the
transverse part of the vacuum propagator is non-zero. However, the
external parameters break manifest Lorentz invariance, hence the gluon
propagator $D$ must be decomposed into chromoelectric and
chromomagnetic modes, $D_E$ and $D_M$, respectively,
\begin{equation}
D_{\mu\nu}(q_0,\vec{q}) = P_{\mu\nu}^{M} D_M(\vec{q}^2,q_0^2) +
 P_{\mu\nu}^{E} D_E(\vec{q}^2,q_0^2)\,.
\label{eq:gluon_decomposition}
\end{equation}
The projectors on the longitudinal and
transversal spatial subspaces, $P_{\mu \nu}^{E}$ and $P_{\mu\nu}^{M}$,
are defined by 
\begin{align}
P_{\mu \nu}^{M} (\vec{q\,},q_0)
 &=  \left(1-\delta_{0\mu} \right)\left(1-\delta_{0\nu} \right)
    \left(\delta_{\mu\nu} -\frac{q_\mu q_\nu}{\vec{q\,}^2} \right)\,,\nonumber\\
P_{\mu \nu}^{E}(\vec{q\,},q_0)
 &= \left(\delta_{\mu\nu}-\frac{q_\mu q_\nu}{q^2} \right)
    -P_{\mu \nu}^{M} (\vec{q\,},q_0)\,.
\label{eq:projectors}
\end{align}

In this section we extend the results presented in \ctrans\ to a wider
area of the $(\mu,T)$ plane.
We have fixed our gauge configurations to the minimal Landau gauge
using the standard overrelaxation algorithm.  The Landau gauge
condition has been imposed with a precision $|\partial_\mu
A_\mu|<10^{-10}$.

In figure~\ref{fig:gluon-compare} we show the two lowest Matsubara
modes for selected spatial momenta as a function of chemical potential
for $N_\tau=24,16,12,8$.  The results shown are for $ja=0.04$, but we
have found no significant difference for $ja=0.02$. We have
investigated the volume dependence 
on the $N_\tau=24$ lattices and found it to be very mild \ctrans.  
At the three lower temperatures, both the electric and magnetic form
factors are roughly independent of $\mu$ up to $\mu a\approx0.5$, and
become suppressed for large $\mu$.  This changes dramatically at the
highest temperature shown ($N_\tau=8$), where for the lowest (static)
Matsubara mode the electric form factor becomes strongly suppressed
with increasing $\mu$, while the magnetic form factor for small
spatial momenta has a clear enhancement at intermedate $\mu$ and an
enhancement at large $\mu$ for larger spatial momenta.  On closer
inspection it is possible to see the onset of this behaviour also for
$N_\tau=12$.  No qualitative differences are seen between the electric
and magnetic form factors for the first nonzero Matsubara mode.

\begin{figure}[tbh]
\begin{center}
\includegraphics*[width=\textwidth]{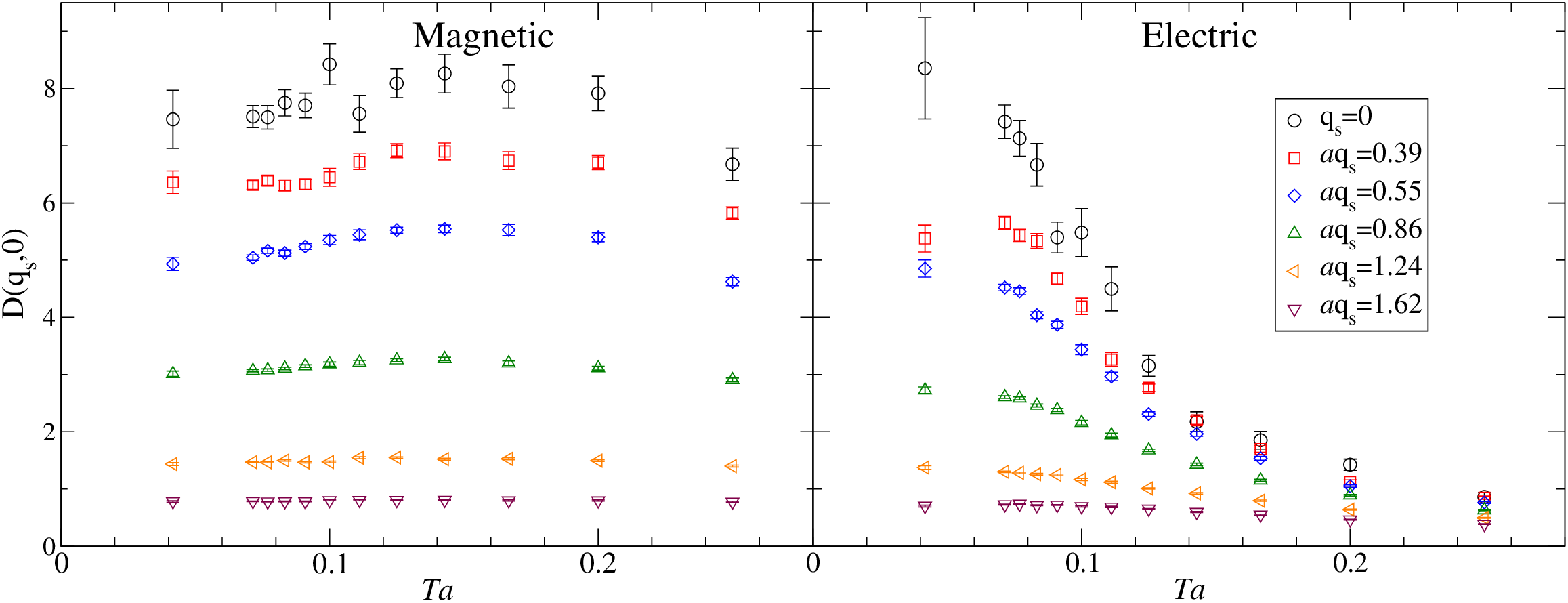}
\caption{Thermal behaviour of the zeroth Matsubara mode of the
  magnetic (left) and electric (right) propagators at $\mu a=0.5$ and
  $ja=0.04$ on $16^3\times N_\tau$ lattices, for selected spatial
  momenta $q_s=|\vec{q}|$.}
\label{fig:zeromodes_vac}
\end{center}
\end{figure}

We now turn to the thermal behaviour of the gluon propagator at fixed
chemical potential.  Fig.\ \ref{fig:zeromodes_vac} shows the zeroth
Matsubara modes of the propagators for $\mu a=0.5$ and $ja=0.04$ on
$16^3\times N_\tau$ lattices as a function of temperature. The
magnetic component has a very mild enhancement at intermediate
temperatures and a slight suppression at very high $T$.  In contrast,
the electric propagator shows a strong suppression with increasing
temperature.  We note that the deconfinement crossover for this value
of $\mu$ happens for $0.8\lesssim Ta\lesssim2.0$, and that this
coincides roughly with the region where the magnetic propagator is
enhanced.  In contrast to early studies in pure Yang--Mills theory,
but in line with a recent study in QCD with twisted-mass Wilson
fermions \cite{Aouane:2012bk}, there is no enhancement in the electric
mode in the transition region.

\section{Summary and outlook}

We have studied the superfluid and deconfinement transition lines in
QC$_2$D in the $(\mu,T)$ plane.  We find that the superfluid
transition temperature is remarkably insensitive to $\mu$ for the
quark mass we are using, while the deconfinement temperature is
clearly decreasing as $\mu$ increases.

At low temperature, the low-momentum modes of both the electric and
magnetic Landau-gauge gluon propagator become suppressed relative to
the (already infrared suppressed) vacuum propagator at large $\mu$,
with no qualitative differences between the two form factors found.
At high temperature, the static electric and magnetic propagators are
found to exhibit very different behaviours, with a strong suppression
of the electric form factor and an enhancement of the magnetic form
factor at intermediate $\mu$.

We are in the process of extending these studies to smaller quark
masses as well as finer lattice spacings.  In a forthcoming
publication we will also study the response of the quark propagator to
$\mu$ and $T$.  This will enable us to directly confront the results
from functional methods for these quantities.

\section*{Acknowledgments}

This work has been
carried out with the support of Science Foundation Ireland grant
11-RFP.1-PHY3193.  We acknowledge the use of the computational
resources provided by the UKQCD collaboration and the
DiRAC Facility jointly funded by STFC, the Large Facilities Capital
Fund of BIS and Swansea University.  We thank the DEISA Consortium
(www.deisa.eu), funded through the EU FP7 project RI-222919, for
support within the DEISA Extreme Computing Initiative.  The simulation
code was adapted with the help of Edinburgh Parallel Computing Centre
funded by a Software Development Grant from EPSRC.
We thank Pietro Giudice, Simon
Hands and Jan Pawlowski for stimulating discussions and advice.

\bibliography{density,hot,lattice,gluon_prop}

\begin{thebibliography}{1}

\bibitem{Hands:2006ve}
S.~Hands, S.~Kim and J.-I. Skullerud,
\newblock Eur. Phys. J. {\bf C48}, 193 (2006), [hep-lat/0604004].

\bibitem{Hands:2010gd}
S.~Hands, S.~Kim and J.-I. Skullerud,
\newblock Phys. Rev. {\bf D81}, 091502R (2010), [arXiv:1001.1682].

\bibitem{Cotter:2012mb}
S.~Cotter, P.~Giudice, S.~Hands and J.-I. Skullerud,
\newblock Phys.Rev. {\bf D87}, 034507 (2013), [arXiv:1210.4496].

\bibitem{Boz:2013rca}
T.~Boz, S.~Cotter, L.~Fister, D.~Mehta and J.-I. Skullerud,
\newblock Eur.Phys.J. {\bf A49}, 87 (2013), [arXiv:1303.3223].

\bibitem{Cotter:2013lat}
S.~Cotter, P.~Giudice, S.~Hands and J.-I. Skullerud,
\newblock PoS {\bf LAT2013} (2013),
\newblock these proceedings.

\bibitem{Aouane:2012bk}
R.~Aouane, F.~Burger, E.-M. Ilgenfritz, M.~Muller-Preussker and A.~Sternbeck,
\newblock Phys. Rev. D 87, {\bf 114502} (2013), [arXiv:1212.1102].

\end{thebibliography}
\end{document}